\documentclass[journal]{IEEEtran}

\ifCLASSINFOpdf

\else

\fi

\usepackage{mathrsfs}
\usepackage{amsmath}
\usepackage{amsfonts}
\usepackage{amsthm}
\usepackage{amssymb}
\usepackage{graphicx}
\usepackage{indentfirst}
\usepackage{array}
\usepackage{cite}
\usepackage[linesnumbered,ruled,vlined]{algorithm2e}
\usepackage{multirow}
\usepackage{latexsym, bm}
\usepackage{booktabs}
\usepackage{color,xcolor}
\usepackage{epstopdf}

\begin{document}
%
\title{\huge Approaching the Finite Blocklength Capacity within 0.025dB by Short Polar Codes and CRC-Aided Hybrid Decoding}
\author{Jinnan Piao, Kai Niu, Jincheng Dai, and Chao Dong
\thanks{This work is supported by National Key R\&D Program of China (No. 2018YFE0205501), 
the National Natural Science Foundation of China (No. 61671080) and BUPT Excellent Ph.D. Students Foundation (No. CX2019218).}
\thanks{The authors are with the Key Laboratory of Universal Wireless Communications, Ministry of Education, Beijing University of Posts and Telecommunications (BUPT), Beijing 100876, China (email: piaojinnan@bupt.edu.cn, niukai@bupt.edu.cn, daijincheng@bupt.edu.cn,  dongchao@bupt.edu.cn).}
\vspace{-2em}
}

\maketitle

\begin{abstract}

In this letter, we explore the performance limits of short polar codes and find that the maximum likelihood (ML) performance of a simple CRC-polar concatenated scheme can approach the finite blocklength capacity. Then, in order to approach the ML performance with a low average complexity, a CRC-aided hybrid decoding (CA-HD) algorithm is proposed and its decoding process is divided into two steps. In the first step, the received sequence is decoded by the adaptive successive cancellation list (ADSCL) decoding. In the second step, CRC-aided sphere decoding with a reasonable initial radius is used to decode the received sequence. To obtain the reasonable radius, the CRC bits of the survival paths in ADSCL are recalculated and the minimum Euclidean distance between the survival path and the received sequence is chosen as the initial radius.
The simulation results show that CA-HD can achieve within about $0.025$dB of the finite blocklength capacity at the block error ratio $10^{-3}$ with code length $128$ and code rate $1/2$. 

\end{abstract}


\IEEEpeerreviewmaketitle

\section{Introduction}

\IEEEPARstart{P}{olar} codes have been proved to achieve the capacity by the successive cancellation (SC) decoding as the code length goes to infinity \cite{arikan}. However, when the code length is small or medium, the performance is unsatisfying. Thus, successive cancellation list (SCL) decoding \cite{talvardyscl, niuscl} is introduced to improve the performance of polar codes. Furthermore, the performance is improved by the CRC-aided SCL (CA-SCL) decoding \cite{niu_CASCL}, which introduces the CRC detector into the SCL decoding.
Based on the CA-SCL, Li \emph{et al.} propose adaptive SCL (ADSCL) decoding which makes a trade-off between performance and complexity \cite{ADSCL}.

Although the list decoding of polar codes has a satisfying performance, it cannot achieve the maximum likelihood (ML) performance with small or medium list size. To reach the ML performance, sphere decoding (SD) algorithm of polar codes is first proposed in \cite{SD} with a cubic complexity \cite{complexity_SD}. Then, the CRC codes are considered into SD and CRC-aided SD (CA-SD) algorithm \cite{CASD} is proposed to achieve the ML performance of the CRC-polar concatenated codes.
However, even though some methods, such as optimizing the path metric \cite{niu_SD} and the radius constraint \cite{efficient_SD}, are used to reduce the complexity, it is still high for short code length. Then, the List-SD algorithm \cite{LSD} proposed with the idea of SCL decoding is also a trade-off between performance and complexity.

Furthermore, to improve the ML performance and approach the finite blocklength capacity, polarization-adjusted convolutional (PAC) codes with the Reed-Muller design rule are proposed by \cite{Arikan_PAC}, but the code rate is fixed as $1/2$.

In this letter, the performance limits of short polar codes are explored. Interestingly, we find that the ML performance of a simple CRC-polar concatenated scheme can approach the finite blocklength capacity.
Then, a CRC-aided hybrid decoding (CA-HD) algorithm for short CRC-polar concatenated codes is proposed so as to approach ML decoding with a low average complexity.
The decoding process of CA-HD is divided into two steps. In the first step, ADSCL is used to decode the received sequence.
In the second step, when the ADSCL decoding is failed and the maximum list size is reached, CA-SD with a reasonable initial radius is started to correct the decoding errors.
In order to facilitate the decoder switching from ADSCL to CA-SD,
the CRC bits of the survival paths in ADSCL are first recalculated to make the survival paths passing the CRC detection. Then, the initial radius of CA-SD is set to the minimum Euclidean distance between the CRC-recalculated survival path and the received sequence.

Moreover, since most of the received sequences are decoded correctly in the first step and the reasonable initial radius can reduce the average decoding complexity of CA-SD in the second step, CA-HD has a low average decoding complexity and is useful to evaluate the ML performance.
The simulation results show that the average complexity of the CA-HD is close to that of the ADSCL with short code length as the signal-to-noise ratio (SNR) increases.
Particularly, we find that the CRC-polar concatenated codes with code length $128$ and code rate $1/2$ can achieve within about $0.025$dB of the finite blocklength capacity at the block error ratio (BLER) $10^{-3}$.

The remainder of the paper is organized as follows. Section II describes the preliminaries of CRC-polar concatenated codes and gives a description of SCL decoding, ADSCL decoding, and CA-SD. Then, the CA-HD is addressed in Section III. Section IV provides the simulation results of different decoding algorithms and presents the decoding complexity comparison. Finally, Section V concludes the letter.

\emph{Notation Conventions}: In this letter, scalars are denoted by the lowercase letters (e.g., $x$). The bold lowercase letters (e.g., ${\bf{x}}$) are used to denote vectors. Notation ${{\bf x}_i^j}$ denotes the subvector $(x_i,\cdots,x_j)$ and $x_i$ denotes the $i$-th element of ${\bf{x}}$.
The calligraphic characters, such as ${\cal X}$, are used to denote sets. The bold capital letters, such as $\mathbf{X}$, denote matrices and the element in the $i$-th row and the $j$-th column of the matrix $\mathbf{X}$ is written as $X_{i,j}$. Throughout this letter, $\bf 1$ means an all-one vector and $\log \left(  \cdot  \right)$ means ``logarithm to base 2''.

\section{Preliminaries}

\subsection{CRC-Polar Concatenated Codes}

For an $(N,K_I+K_C)$ CRC-polar concatenated code with code length $N = 2^n$ and code rate $R = K_I/N$, the message sequence $\bf b$ is first encoded by a $(K, K_I)$ CRC code, and the CRC bits are attached to the end of message sequence, where $K_I$ is the length of message sequence, $K_C$ is the length of CRC bits and $K$ is $K_I+K_C$. Given the CRC generator polynomial $g(x)$, the polynomial of the CRC encoded sequence $\bf s$ is denoted by
\begin{equation}\label{CRC_encoding}
s(x) = x^{K_C} \cdot b(x) + \left(x^{K_C}\cdot b(x)\right) {\rm mod}~g(x),
\end{equation}
where $b(x)$ is the polynomial of message sequence $\bf b$.

Then, $\bf s$ is treated as the information bits of an $(N,K)$ polar code and inserted into the information sequence $\bf u$ in terms of the information set $\cal A$, where $\bf u$ is an $N$-length sequence and $u_i$ is assigned to information bit if $i \in {\cal A}$, or 0 if $i \in {\cal A}^c$. Furthermore, a codeword ${\bf c}$ of the polar code is calculated by ${\bf c} = {\bf uB}{\bf G} = {\bf v}{\bf G}$, where $\bf B$ is a bit-reversal permutation matrix, $\bf v$ is ${\bf uB}$, ${\bf G}$ is an $n$-th Kronecker power of ${\bf G}_2$ and ${\bf G}_2$ is ${\begin{bmatrix}
\begin{smallmatrix}
1&0\\
1&1
\end{smallmatrix}
\end{bmatrix}}$.

Without loss of generality, the BI-AWGN channel and BPSK modulation are considered in this letter. Thus, each coded bit $c_i$ is modulate into the transmitted signal by $x_i = 1 - 2c_i$. Then, the received sequence is ${\bf y} = {\bf x} + {\bf n}$, where $n_i$ is i.i.d. AWGN with zero mean and variance $\sigma^2$. Thus, the SNR is defined as $\frac{E_s}{N_0}$, i.e.,
\begin{equation}
    \frac{E_s}{N_0} = \frac{1}{2\sigma^2},
\end{equation}
where $E_s$ is the energy of the transmitted signal and $N_0$ is the one-sided power spectral density of the noise.

\subsection{SCL Decoding and ADSCL Decoding}

Polar codes can be decoded by the SC decoding algorithm with the decoding complexity $O(N\log N)$.
Although SC decoding can achieve the channel capacity with the infinite code length, its performance is unsatisfied for the short or medium code length.

Thus, the SCL decoding is proposed and its decoding process is similar to the SC decoding except keeping at most $L$ survival paths. When an information bit $u_{i-1}$ is decoded, the $l$-th survival path is split into two paths with attempting ${\hat u}_{i_l} = 0$ and ${\hat u}_{i_l} = 1$, where ${\hat u}_{i_l}$ is the $i$-th decoded bit of the $l$-th survival path. Then, when the number of paths is larger than $L$, the best $L$ paths are reserved and other paths are discarded.
Furthermore, by introducing the CRC detector and the maximum list size $L_{\max}$,
ADSCL is proposed to make a trade-off between performance and complexity.

\subsection{CA-SD Algorithm}

ML decoding of CRC-polar concatenated codes is equivalent to the minimization problem as follows
\begin{equation}\label{min_problem}
\begin{aligned}
{\hat {\bf{v}}} = \mathop {\arg \min }\limits_{{\bf{x}} } {\left\| {{\bf{y}} - {\bf{x}}} \right\|^2} = \mathop {\arg \min }\limits_{{\bf{v}} } {\left\| {{\bf y} - \left({\bf 1} - 2{\bf v}{\bf G}\right)} \right\|^2}.
\end{aligned}
\end{equation}
CA-SD can solve (\ref{min_problem}) by means of enumerating the sequence $\bf v$ satisfying the constraint
\begin{equation}\label{constraint}
  d\left({\bf v}_1^N\right) \triangleq {\left\| {{\bf y} - \left({\bf 1} - 2{\bf v}{\bf G}\right)} \right\|^2} \le r^2, {\bf v} \in {\cal V},
\end{equation}
where the elements of the set ${\cal V}$ can pass the CRC detection
and $r$ is the radius for the CA-SD.
To make sure ${\bf v} \in {\cal V}$, the bits of $\bf v$ are divided into information bits, frozen bits and parity check bits, where parity check bits are directly computed by the previous decoded bits to avoid the redundant search.

\section{CA-HD Algorithm}

In this section, we first introduce the decoder structure of CA-HD. Then, the details of CA-HD are described.

\subsection{Decoder Structure}

\begin{figure}[h]
\setlength{\abovecaptionskip}{0.cm}
\setlength{\belowcaptionskip}{-0.cm}
  \centering{\includegraphics[scale=1.2]{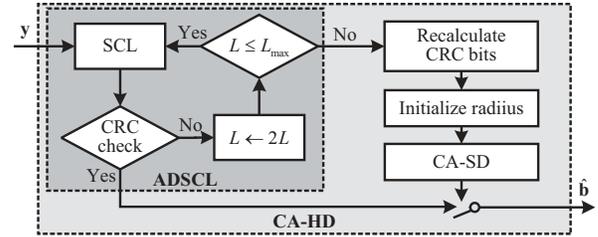}}
  \caption{The block diagram of the decoder structure of CA-HD.}\label{decoder_structure}
  \vspace{-1em}
\end{figure}

The decoder structure of CA-HD is illustrated in Fig. \ref{decoder_structure}. In Fig. \ref{decoder_structure}, the received sequence $\bf y$ is first decoded by the ADSCL with the maximum list size $L_{\max}$. During the ADSCL, when a survival path can pass the CRC detection, this path is regarded as the decoded sequence and the CA-HD is finished. Otherwise, if $L_{\max}$ is reached and no paths can pass the CRC detection, CA-SD is used to decode $\bf y$, where the initial radius $r_0$ is calculated by the $L_{\max}$ survival paths.

\subsection{Detailed Description}

Algorithm \ref{CAHD} describes the entire procedure of CA-HD. In Algorithm \ref{CAHD}, the processes of ADSCL and CA-SD are similar to those in \cite{ADSCL} and \cite{CASD}, respectively.

For the ADSCL, it begins from the SCL with list size $L = 1$. Then, if no survival paths can pass the CRC detection, $L$ is set to $2L$, and the received sequence is decoded by SCL continuously until passing CRC detection or $L$ larger than $L_{\max}$. The path metric in the ADSCL based on log likelihood ratio (LLR) \cite{LLRSCL} is computed as
\begin{equation}
\begin{array}{l}
{\rm PM}_{ - {1_l}} = 0,\\
{\rm PM}_{i_l} = \left\{ \begin{array}{l}
{\rm PM}_{{{\left( {i - 1} \right)}_l}} + \left| {{\alpha _{{i_l}}}} \right|,~{\rm if}~{{\hat u}_{{i_l}}} \ne \frac{1}{2}\left( {1 - {\mathop{\rm sgn}} \left( {{\alpha _{{i_l}}}} \right)} \right),\\
{\rm PM}_{{{\left( {i - 1} \right)}_l}}~~~~~~~~~,~{\rm otherwise},
\end{array} \right.
\end{array}
\end{equation}
where $\alpha _{i_l}$ is the LLR of $u_{i_l}$.

For the CA-SD, CRC is used to improve the distance spectrum of polar codes rather than detecting error. Then, since ${\bf G}$ is an $n$-th Kronecker power of ${\bf G}_2$, it is a lower triangular matrix. Thus, the partial square-Euclidean distance along with ${\bf v}_i^N$ is defined as
\begin{equation}
  \begin{array}{l}
  d\left({\bf v}_i^N\right) \triangleq \sum\limits_{k = i}^N {\left| {{y_k} - \left(1-2\cdot{\mathop  \oplus \limits_{j=k}^N \left(G_{j,k}  v_j\right)}\right)} \right|}^2,
  \end{array}
\end{equation}
which is recursively calculated as
\begin{equation}\label{SD_euclidean_distance}
    d\left({\bf v}_i^N\right) = d\left({\bf v}_{i+1}^N\right) + {\left| {{y_i} - \left(1-2\cdot{\mathop  \oplus \limits_{j=i}^N \left(G_{j,i} v_j\right)}\right)} \right|}^2,
\end{equation}
where `$\oplus$' denotes summation over $GF(2)$. Therefore, according to (\ref{SD_euclidean_distance}), the search order of CA-SD is from ${\bf v}_N$ to ${\bf v}_1$. When a valid sequence ${\bf v}$ satisfying (\ref{constraint}) is found, the radius $r$ is updated with $\sqrt{d\left({\bf v}_1^N\right)}$ and CA-SD continues until we find the ML solution.

The process of optimizing the initial radius is described in line $8$ to line $13$ of Algorithm \ref{CAHD}.
A reasonable initial radius should make sure that at least one information sequence passing the CRC detection can satisfy the radius constraint of the CA-SD, i.e,
\begin{equation}\label{reasonable_initial_radius}
    \exists {\bf{u}} \in {\cal U}, \left\| {\bf y} - \left({\bf 1} - 2{{\bf u}}{\bf B}{\bf G}\right)  \right\| \le r_0,
\end{equation}
where the elements of the set $\cal U$ can pass the CRC detection and $r_0$ is an initial radius.
However, since the $L_{\max}$ survival paths of ADSCL cannot pass the CRC detection when ADSCL is failed, the initial radius calculated by the survival paths directly is unreasonable.
This means that if the initial radius $r_0$ is assigned based on the survival path, there will be no information sequence satisfying the radius constraint (\ref{reasonable_initial_radius}). Thus, using this assignment, CA-SD cannot find the ML sequence.

In order to obtain a reasonable initial radius, $\hat{\bf u}_l$ should be modified to make $\hat{\bf u}_l \in {\cal U}$. A simple method is to revise the CRC bits of $\hat{\bf u}_l$. To calculate the new CRC bits, we first obtain the decoded message sequence ${\hat{\bf b}}_l$ from ${\hat{\bf u}}_l$ according to $\cal A$. Then, ${\hat{\bf b}}_l$ is encoded into ${\hat{\bf s}}_l$ by (\ref{CRC_encoding}) and the new CRC bits are the last $K_C$ bits of ${\hat{\bf s}}_l$. Next, the CRC bits of $\hat{\bf u}_l$ are replaced by these new CRC bits and the newly survival path is denoted by ${\tilde{\bf u}}_l$. Due to ${\tilde{\bf u}}_l \in {\cal U}$, the initial radius calculated by it is reasonable.
Thus, the optimized initial radius of CA-SD is
\begin{equation}\label{radius}
    r_0 = \mathop {\min }\limits_l  \left\| {\bf y} - \left({\bf 1} - 2{\tilde{\bf u}}_l{\bf B}{\bf G}  \right\|\right), l = 1,2,\cdots,L_{\max}.
\end{equation}

\begin{algorithm}[t]
\setlength{\abovecaptionskip}{0.cm}
\setlength{\belowcaptionskip}{-0.cm}
\caption{The proposed CA-HD algorithm}\label{CAHD}
\KwIn {The received sequence $\bf y$;}
\KwOut {The decoded message sequence $\hat{\bf b}$\;}
Initialize the list size $L \leftarrow 1$\;
\While{$L \le L_{\max}$}
{
    $\bf y$ is decoded by SCL with list size $L$\;
    \For{$l = 1 \to L$}
    {
        \If{${\hat{\bf u}}_l$ can pass the CRC detection}
        {
            Obtain $\hat{\bf b}$ from ${\hat{\bf u}}_l$ and go to line 15\;
        }
    }
    $L \leftarrow 2L$\;
}
\For{$l = 1 \to L_{\max}$}
{
    According to $\cal A$, obtain ${\hat{\bf b}}_l$ from ${\hat{\bf u}}_l$\;
    ${\hat{\bf b}}_l$ is encoded into ${\hat{\bf s}}_l$  by (\ref{CRC_encoding})\;
    Replace the CRC bits of ${\hat{\bf u}}_l$ by the last $K_C$ bits of ${\hat{\bf s}}_l$ and the newly survival path is denoted by ${\tilde{\bf u}}_l$\;
    A reasonable radius is $\left\| {\bf y} - \left({\bf 1} - 2{\tilde{\bf u}}_l{\bf B}{\bf G}\right)  \right\|$\;
}
Set the initial radius $r_0$ by (\ref{radius})\;
$\bf y$ is decoded by CA-SD with $r_0$ and obtain $\hat{\bf b}$\;
$\hat{\bf b}$ is the decoded message sequence\;
\end{algorithm}

\section{Performance Evaluation}

In this section,  the BLER performance and the complexities of CA-HD, CA-SD, CA-SCL, and ADSCL are provided.
We also provide the normal approximation of the finite blocklength capacity of BI-AWGN channel \cite{NormalApproximation}, i.e.,
\begin{equation}
    R \approx C - \sqrt{\frac{V}{N}}Q^{-1}(P_e)+\frac{\log N}{2N},
\end{equation}
where $C$ is the channel capacity, $V$ is the channel dispersion and $P_e$ is the error probability.
The Gaussian approximation (GA) \cite{GA_Trifonov} is applied to construct polar codes.
The maximum list size $L_{\max}$ in ADSCL decoding and CA-HD is set as $1024$.
The list sizes of CA-SCL are 8 and 32

Table \ref{Table_CRC_polynomial} provides the optimized CRC polynomials for short polar codes with code length $N$, code rate $R$ and CRC length $K_C$, where $g(x) = g_{K_C}x^{K_C}+ \cdots + g_1x + g_0$ is denoted as a vector $(g_{K_C}\cdots g_1g_0)$, $d_{\min}$ is the minimum Hamming weight and $A_{d_{\min}}$ is the number of the codewords with $d_{\min}$.
The CRC polynomial is optimized by brute-force searching $d_{\min}$ and $A_{d_{\min}}$ of the CRC-polar concatenated codes.

\begin{table}[t]
\centering
\caption{The Optimized CRC Polynomials for Short Polar Codes}
\label{Table_CRC_polynomial}
\begin{tabular}{p{0.2cm}<{\centering}p{0.3cm}<{\centering}p{0.3cm}<{\centering}ccc}
  \toprule
  $N$ & $R$ & $K_C$ & $g(x)$ & $d_{\min}$ & $A_{d_{\min}}$\tabularnewline
  \midrule
  \multirow{3}{*}{$64$} & $1/3$ & $12$ & $(1100110100101)$ & 16 & 168\tabularnewline
  & $1/2$ & $13$ & $(11110101010101)$ & 10 & 34 \tabularnewline
  & $2/3$ & $18$ & $(1010110011010001001)$ & 8 & 4238 \tabularnewline
  \midrule
  \multirow{3}{*}{$128$} & $1/3$ & $20$ & $(100000000010111010001)$ & 24 & 171  \tabularnewline
  & $1/2$ & $24$ & $(1000000000000000111100101)$ & 16 & 66 \tabularnewline
  & $2/3$ & $16$ & $(10001011110110111)$ & 10 & 167  \tabularnewline
  \bottomrule
\end{tabular}
\vspace{-1em}
\end{table}

\begin{figure}[t]
\setlength{\abovecaptionskip}{0.cm}
\setlength{\belowcaptionskip}{-0.cm}
  \centering{\includegraphics[scale=0.62]{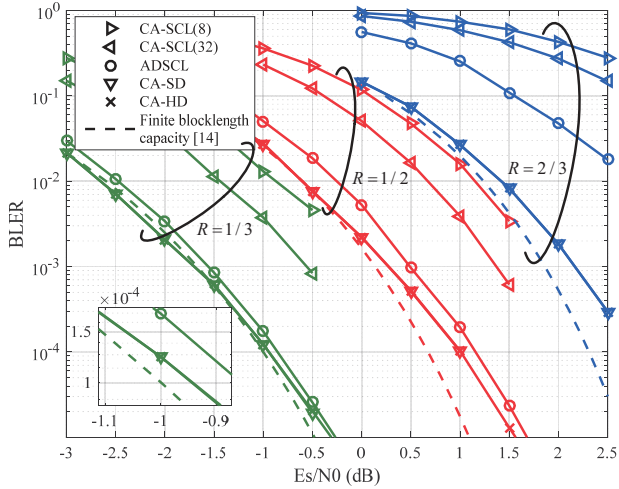}}
  \caption{The BLER performance of CA-HD, CA-SD, CA-SCL and ADSCL with code length $64$ and code rates $1/3$, $1/2$ and $2/3$.}\label{BLER64}
  \vspace{-1em}
\end{figure}

Fig. \ref{BLER64} illustrates the BLER performance of CA-HD, CA-SD, CA-SCL and ADSCL with code length $64$ and code rates $1/3$, $1/2$ and $2/3$.
For the three code rates, the performance of CA-HD is better than that of ADSCL and CASCL. In addition, it coincides with the performance of CA-SD, which means the proposed CA-HD can approach the ML performance.
Moreover, when $R$ is $1/3$, the performance of CA-HD can approach the finite blocklength capacity.
When $R$ is $1/2$, the BLER curve of CA-HD is close to the finite blocklength capacity in the low to medium SNR regions and the performance gap between the two curves becomes larger with the increase of SNR. When $R$ is $2/3$, the performance of CA-HD is close to the finite blocklength capacity in the low SNR region only.

\begin{figure}[t]
\setlength{\abovecaptionskip}{0.cm}
\setlength{\belowcaptionskip}{-0.cm}
  \centering{\includegraphics[scale=0.62]{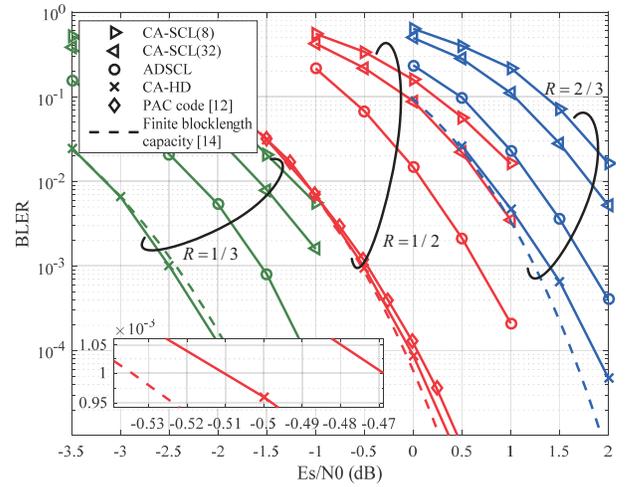}}
  \caption{The BLER performance of CA-HD, CA-SCL and ADSCL with code length 128 code rates $1/3$, $1/2$ and $2/3$.}\label{BLER128}
\end{figure}

Fig. \ref{BLER128} provides the BLER performance with $N = 128$ and the three code rates. For the three code rates, the trend of CA-HD performance with $N = 128$ is similar to that with $N = 64$. Then, the comparison between CA-HD and PAC codes \cite{Arikan_PAC} with $R = 1/2$ is also provided. The performance of the CA-HD is almost the same as that of and the PAC code in the low to medium SNR region, but the performance gap between the two curves becomes larger as the SNR increases and the performance of CA-HD is better in the high SNR region. Specifically, at the BLER $10^{-3}$ and $10^{-4}$, the performance of the CA-HD is better than that of the PAC code.

Interestingly, when $N$ is $64$ and $R$ is $1/3$, the performance gap between the CA-HD and the finite blocklength capacity is about $0.05$dB at the BLER $10^{-4}$. Further, we find that the CA-HD with $N = 128$ and $R = 1/2$ can achieve within about $0.025$dB of the finite blocklength capacity at the BLER $10^{-3}$. To the best of our knowledge, this result is the first example close to the finite blocklength capacity so much.

\begin{figure}[t]
\setlength{\abovecaptionskip}{0.cm}
\setlength{\belowcaptionskip}{-0.cm}
  \centering{\includegraphics[scale=0.62]{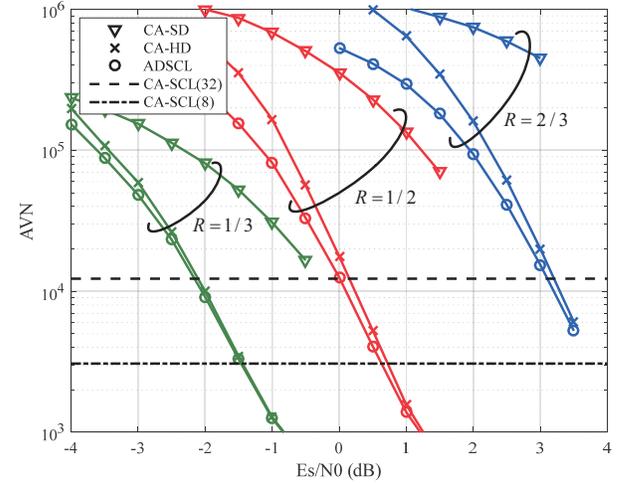}}
  \caption{The complexity comparison among CA-HD, CA-SD, CA-SCL and ADSCL with code length 64 and code rates $1/3$, $1/2$ and $2/3$.}\label{Complexity64}
  \vspace{-1em}
\end{figure}

Fig. \ref{Complexity64} shows the complexities of the four decoding algorithms with $N = 64$ and the three code rates. The complexity is counted by the average visited nodes (AVN).
The AVN of ADSCL and CA-SCL are counted as ${\bar L}N\log N$ and $LN\log N$, respectively,
where ${\bar L}$ is the average list size of ADSLC and $L = 32$ or $8$ is the list size of CA-SCL.
It is observed that for the three code rates, the complexity of CA-HD is located between the complexities of ADSCL and CA-SD.
Then, it approaches the complexity of ADSCL as the SNR increases and is lower than that of CA-SCL in the high SNR region.
Moreover, although CA-HD and CA-SD both achieve the ML performance, the average complexity of CA-HD is up to $100$ times lower than that of CA-SD.
The reasons are that most of the received sequences can be decoded correctly in the first step and the reasonable initial radius can reduce the average decoding complexity of CA-SD in the second step.
Thus, the average complexity of CA-HD is close to that of ADSCL and lower than that of CA-SCL, which makes CA-HD useful to evaluate the ML performance of short polar codes.

\section{Conclusion}

In this letter, we explore the performance limits of short polar codes and propose a CA-HD algorithm to approach the ML performance.
The performance of CA-HD coincides with that of CA-SD and a lower average complexity. Thus, it is useful to evaluate the ML performance of short polar codes.
The simulation results show that CA-HD can achieve within about $0.025$dB of the finite blocklength capacity at BLER $10^{-3}$ with code length $128$ and code rate $1/2$.

\vspace{-1em}

\begin{thebibliography}{99}

\bibitem{arikan}
E. Ar{\i}kan, ``Channel polarization: A method for constructing capacity achieving codes for symmetric binary-input memoryless channels,'' \emph{IEEE Trans. Inf. Theory}, vol. 55, no. 7, pp. 3051--3073, Jul. 2009.

\bibitem{talvardyscl}
I. Tal and A. Vardy, ``List decoding of polar codes,'' \emph{IEEE Trans. Inf. Theory}, vol. 61, no. 5, pp. 2213--2226, May. 2015.

\bibitem{niuscl}
K. Chen, K. Niu, and J. R. Lin, ``List successive cancellation decoding
of polar codes,'' \emph{Electron. Lett.}, vol. 48, no. 9, pp. 500--501, 2012.

\bibitem{niu_CASCL}
K. Niu and K. Chen, ``CRC-aided decoding of polar codes,'' \emph{IEEE Commun. Lett.}, vol. 16, no. 10, pp. 1668--1671, Oct. 2012.

\bibitem{ADSCL}
B. Li, H. Shen and D. Tse, ``An Adaptive Successive Cancellation List Decoder for Polar Codes with Cyclic Redundancy Check,'' \emph{IEEE Commun. Lett.}, vol. 16, no. 12, pp. 2044--2047, December 2012.

\bibitem{SD}
S. Kahraman and M. E. Celebi, ``Code based efficient maximum-likelihood decoding of short polar codes,'' in \emph{Proc. 2012 IEEE Int. Symp. Inf. Theory}, Cambridge, MA, pp. 1967--1971, 2012.

\bibitem{complexity_SD}
B. Hassibi and H. Vikalo, ``On the sphere-decoding algorithm I. expected complexity,'' \emph{IEEE Trans. Signal Processing}, vol. 53, no. 8, pp. 2806--2818, Aug. 2005.

\bibitem{CASD}
J. Piao, J. Dai and K. Niu, ``CRC-Aided Sphere Decoding for Short Polar Codes,'' \emph{IEEE Commun. Lett.}, vol. 23, no. 2, pp. 210--213, Feb. 2019.

\bibitem{niu_SD}
K. Niu, K. Chen and J. Lin, ``Low-Complexity Sphere Decoding of Polar Codes Based on Optimum Path Metric,'' \emph{IEEE Commun. Lett.}, vol. 18, no. 2, pp. 332--335, Feb. 2014.

\bibitem{efficient_SD}
J. Guo and A. Guill¨¦n i F¨¤bregas, ``Efficient sphere decoding of polar codes,'' in \emph{Proc. 2015 IEEE Int. Symp. Inf. Theory}, Hong Kong, pp. 236--240, 2015.

\bibitem{LSD}
S. A. Hashemi, C. Condo and W. J. Gross, ``List sphere decoding of polar codes,'' in \emph{Proc. 2015 49th Asilomar Conference on Signals, Systems and Computers}, Pacific Grove, CA, pp. 1346--1350, 2015.

\bibitem{Arikan_PAC}
E. Ar{\i}kan, ``From sequential decoding to channel polarization and back again,'' \emph{arXiv preprint arXiv:1908.09594 (2019).}

\bibitem{LLRSCL}
A. Balatsoukas-Stimming, M. B. Parizi and A. Burg, ``LLR-Based Successive Cancellation List Decoding of Polar Codes,'' \emph{IEEE Trans. Signal Processing}, vol. 63, no. 19, pp. 5165--5179, Oct.1, 2015.

\bibitem{NormalApproximation}
Y. Polyanskiy, H. V. Poor and S. Verdu, ``Channel Coding Rate in the Finite Blocklength Regime,'' \emph{IEEE Trans. Inf. Theory}, vol. 56, no. 5, pp. 2307--2359, May 2010.

\bibitem{GA_Trifonov}
P. Trifonov, ``Efficient design and decoding of polar codes,'' \emph{IEEE Trans. Commun.}, vol. 60, no. 11, pp. 3221--3227, Nov. 2012.





\end{thebibliography}
\end{document}